\documentclass[compsoc, conference, a4paper, 10pt, times]{IEEEtran}

\usepackage{cite}
\usepackage{amsmath,amssymb,amsfonts}
\usepackage{algorithmic}
\usepackage{graphicx}
\usepackage{textcomp}
\usepackage{xcolor}
\usepackage{booktabs}

\usepackage[hidelinks, pdftex,
            pdfauthor={Romy Fieblinger, Md Tanvirul Alam, Nidhi Rastogi},
            pdftitle={Actionable Cyber Threat Intelligence using Knowledge Graphs and Large Language Models},
            pdfkeywords={Cyber Threat Intelligence, Large Language Models, Knowledge Graphs, Threat Prediction}]{hyperref}
            
\usepackage[shortlabels]{enumitem}
\usepackage{textgreek}
\usepackage{array}
\usepackage{caption}
\usepackage{booktabs}
\usepackage{lipsum}
\usepackage{amsmath,amsfonts,amssymb}

\usepackage{tcolorbox}

\usepackage{booktabs}
\usepackage{multirow}
\usepackage{multicol}
\usepackage{color,soul}
\usepackage{listings}
\usepackage{inconsolata}
\tcbuselibrary{listings, skins}

\definecolor{lightgray}{RGB}{200, 200, 200}
\definecolor{lightergray}{RGB}{230, 230, 230}
\definecolor{darkgray}{RGB}{150, 150, 150}
\definecolor{highlightcolor}{RGB}{0, 150, 200}

\newcommand{\code}[1]{\sethlcolor{lightergray}\texttt{\small\hl{#1}}}

\newtcolorbox{highlighted}[1]{%
    colback=lightgray,
    colframe=darkgray,
    boxrule=0.5pt,
    arc=2pt,
    boxsep=2pt,
    left=3pt,
    right=3pt,
    top=2pt,
    bottom=2pt,
    #1
}

\definecolor{codegreen}{rgb}{0,0.6,0}
\definecolor{codegray}{rgb}{0.5,0.5,0.5}
\definecolor{codepurple}{rgb}{0.58,0,0.82}
\definecolor{backcolour}{rgb}{0.95,0.95,0.92}

\newtcblisting{mycode}{
  listing only,
  boxrule=0pt,
  arc=0pt,
  bottomrule=0pt, 
  top=0mm, 
  bottom=0mm, 
  boxsep=0.005pt, 
  left=5mm,
  right=0pt,
  colback=gray!20, 
  listing options={
    basicstyle=\sffamily\tiny,
    language=Python,
    breaklines=true
  },
}
\lstdefinestyle{mystyle}{
    backgroundcolor=\color{lightergray},   
    commentstyle=\color{codegreen},
    keywordstyle=\color{magenta},
    numberstyle=\tiny\color{codegray},
    stringstyle=\color{codepurple},
    basicstyle=\ttfamily\scriptsize,
    breakatwhitespace=false,         
    breaklines=true,                 
    captionpos=b,                    
    keepspaces=true,                 
    numbers=left,                    
    numbersep=5pt,                  
    showspaces=false,                
    showstringspaces=false,
    showtabs=false,                  
    tabsize=2
}
\newtcblisting{myinst}{
  listing only,
  boxrule=0pt,
  arc=0pt,
  bottomrule=0pt, 
  top=0mm, 
  bottom=0mm, 
  boxsep=0.05pt, 
  left=5mm,
  right=0pt,
  colback=gray!20, 
  listing options={
    basicstyle=\ttfamily\scriptsize\color{black},
    keywordstyle=\color{black},
    language=Python,
    breaklines=true,
  },
}

\usepackage{fancyhdr}

\fancypagestyle{firstpage}{
  \fancyhf{} 
  \fancyfoot[L]{\footnotesize 6th Workshop on Attackers and Cyber-Crime Operations} 
}

\begin{document}

\title{Actionable Cyber Threat Intelligence using Knowledge Graphs and Large Language Models}

\author{\IEEEauthorblockN{1\textsuperscript{st} Romy Fieblinger}
\IEEEauthorblockA{\textit{Mittweida University of Applied Sciences} \\
Mittweida, Germany \\
rfieblin@hs-mitweida.de}
\and
\IEEEauthorblockN{2\textsuperscript{nd} Md Tanvirul Alam}
\IEEEauthorblockA{\textit{Rochester Institute of Technology} \\
Rochester NY, USA\\
ma8235@rit.edu}
\and
\IEEEauthorblockN{3\textsuperscript{rd} Nidhi Rastogi}
\IEEEauthorblockA{\textit{Rochester Institute of Technology} \\
Rochester NY, USA \\
nidhi.rastogi@rit.edu}
}

\maketitle

\thispagestyle{firstpage}

\begin{abstract}
Cyber threats are constantly evolving. Extracting actionable insights from unstructured Cyber Threat Intelligence (CTI) data is essential to guide cybersecurity decisions. Increasingly, organizations like Microsoft, Trend Micro, and CrowdStrike are using generative AI to facilitate CTI extraction. This paper addresses the challenge of automating the extraction of actionable CTI using advancements in Large Language Models (LLMs) and Knowledge Graphs (KGs). We explore the application of state-of-the-art open-source LLMs, including the Llama 2 series, Mistral 7B Instruct, and Zephyr for extracting meaningful triples from CTI texts. Our methodology evaluates techniques such as prompt engineering, the guidance framework, and fine-tuning to optimize information extraction and structuring. The extracted data is then utilized to construct a KG, offering a structured and queryable representation of threat intelligence. Experimental results demonstrate the effectiveness of our approach in extracting relevant information, with guidance and fine-tuning showing superior performance over prompt engineering. However, while our methods prove effective in small-scale tests, applying LLMs to large-scale data for KG construction and Link Prediction presents ongoing challenges.
\end{abstract}


\section{Introduction}
The ever-evolving landscape of cyber threats has made incident threat analysis challenging, even for experienced professionals. Cyber Threat Intelligence (CTI) addresses this by providing information on threat actors, including indicators of compromise (IoCs) such as IP addresses or file hashes, as well as attacker tactics, techniques, and procedures (TTPs). This information is crucial for cybersecurity analysts, enabling them to make informed security decisions and stay updated on new threats, per a 2023 CTI survey by SANS Security~\cite{brown_sans_2023}.
However, CTI is primarily provided in an unstructured format and can be noisy, making knowledge extraction labor-intensive, with over half of the security teams spending more than 40\% of their time on these tasks \cite{brown_sans_2023}. Therefore, we need more automation to reduce the overwhelming volume of open source reporting and the large amount of time spent analyzing it \cite{brown_sans_2023}.
To address these challenges, methodologies like TTPHunter~\cite{rani_ttphunter_2023} and LADDER~\cite{alam_looking_2023} have been proposed to structure CTI reports into Knowledge Graphs (KGs).
Despite their contributions, they often face limitations, such as high false positive rates in classifying IoCs and TTPs, scalability challenges, and the potential to miss critical information due to the constraints of language and predefined schemas.

Lately, large language models (LLMs) have proven valuable for understanding and manipulating natural language. With tailored prompts or additional training, they can perform effectively in tasks such as Named Entity Recognition \cite{wang_gpt-ner_2023}, Relation Extraction \cite{wan_gpt-re_2023}\cite{wadhwa_revisiting_2023}, and Triple Extraction for constructing Knowledge Graphs \cite{zhu_llms_2023}\cite{han_pive_2023}\cite{trajanoska_enhancing_2023}. Cybersecurity applications include interpreting TTPs \cite{fayyazi_uses_2023} and comprehending cybersecurity terminologies \cite{juttner_chatids_2023}\cite{gupta_chatgpt_2023}. However, LLMs' potential for extracting and interpreting critical information from natural language-heavy CTI remains largely unexplored.

\begin{figure*}[ht!]
     \centering
   \resizebox{.95\textwidth}{!}{ \fbox{\includegraphics[width=\textwidth]{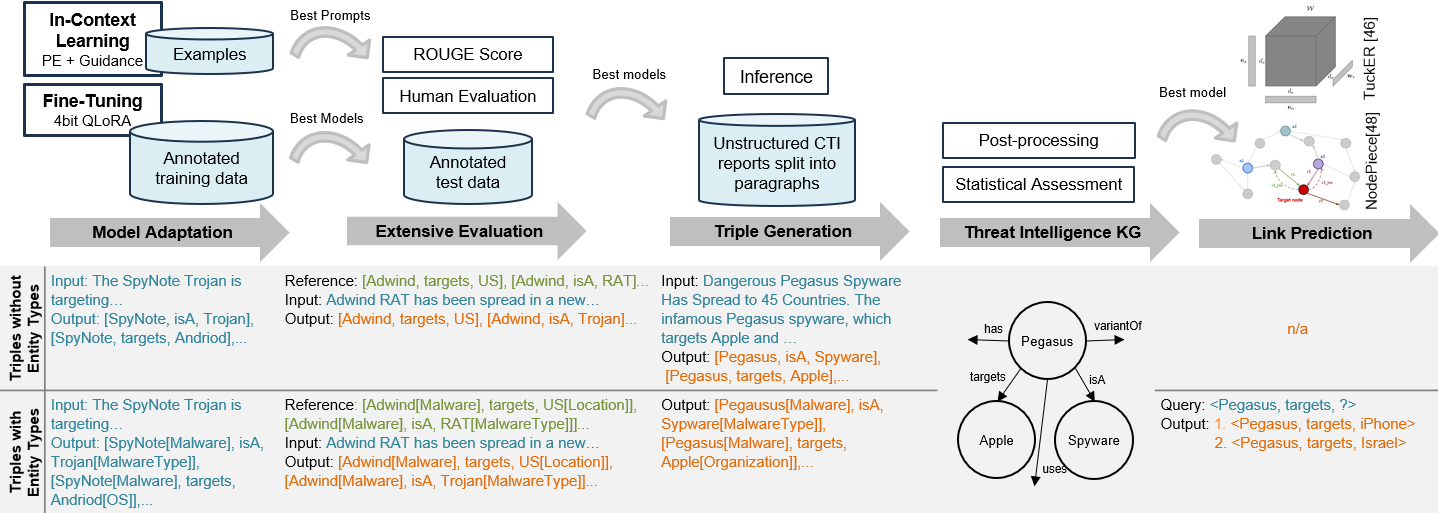}}}
     \captionsetup{font=scriptsize}
     \caption{Proposed approach: Outline for LLM-based CTI extraction and KG development. Initial stages involve adapting the models with Few-Shot Prompting and Fine-tuning to generate triple output. Following this, extensive evaluation determines the best model and prompt combination. The top models are then utilized for triple generation in the KG, leading to Link Prediction on the optimal KG, showcasing the transformation from raw data to actionable intelligence.}
     \label{fig:diagram}
\end{figure*}

Therefore, in this research, we address the challenge of automatically extracting actionable Cyber Threat Intelligence (CTI) using Large Language Models (LLMs) and Knowledge Graphs (KGs) (see Figure \ref{fig:diagram}). Specifically, we investigate the application of open-source LLMs, including the Llama 2 series~\cite{touvron_llama_2023}, Mistral 7B Instruct~\cite{jiang_mistral_2023}, and Zephyr~\cite{tunstall_zephyr_2023}, for extracting triples from CTI text. Our approach includes the evaluation of various techniques like few-shot learning with prompt engineering~\cite{openai_prompt} and the guidance framework~\cite{noauthor_guidance-aiguidance_2024}, as well as fine-tuning~\cite{hu_lora_2021}. The best-performing model, as determined by the ROUGE score and human evaluation, is then employed to generate a KG from CTI reports. This graph is subsequently applied to link prediction tasks, showcasing the practical implications of our work in enhancing cybersecurity defenses and response strategies.

The paper makes the following key contributions:
\begin{enumerate}
\item  We investigate using LLMs to extract information from unstructured CTI for KG construction. While prior work used machine learning, we are the first, to our knowledge, to employ LLMs here.

\item  We explore this task under few-shot prompting and fine-tuning settings, testing various LLMs,  prompts, inference, and decoding strategies. Our study offers practical insights for effective LLM use in this context. 

\item We identify LLM limitations for large-scale datasets and propose practical solutions. By highlighting these challenges, we aim to advance LLM applicability and scalability in information extraction.
\end{enumerate}

\section{Background and Related Work}
\textbf{Cyber Threat Intelligence (CTI).} CTI involves gathering and analyzing security threats information to aid decision-making and help organizations respond effectively \cite{johnson_guide_2016}. 
Public CTI sources include threat reports, vendor blogs (e.g., Symantec~\cite{symantec}, Mandiant~\cite{mandiant}, CrowdStrike~\cite{crowdstrike}), news sites, databases, and social media. Effective CTI must be relevant to the organization, actionable for immediate threat response, and contribute to key business outcomes \cite{dalziel2014define}.
In recent years, several approaches have emerged for extracting information from CTI reports, primarily focusing on TTPs. Early works like TTPDrill~\cite{husari_ttpdrill_2017} and ChainSmith~\cite{zhu_chainsmith_2018} used NLP to extract threat actions and IoCs. 
CASIE~\cite{satyapanich_casie_2020} combined deep learning and linguistic features for cybersecurity event and vulnerability extraction.
More recently, transformer-based architectures like SecureBERT~\cite{aghaei_securebert_2022} and TTPHunter~\cite{rani_ttphunter_2023} fine-tuned models such as BERT~\cite{devlin2019bert} or RoBERTa~\cite{liu2019roberta} for better cybersecurity text classification and context-sensitive TTP extraction.

\textbf{Large Language Models (LLMs).} LLMs are transformer-based models \cite{vaswani_attention_2017} that excel in language modeling by estimating the probability of word sequences and generating text. These models undergo pre-training on extensive textual corpora, significantly enhancing their effectiveness in various NLP tasks, such as text generation and machine translation by providing context-aware representations. LLMs are characterized by their large scale, with tens or hundreds of billions of parameters \cite{zhao_survey_2023}. \textbf{Fine-tuning} is commonly applied to models like Llama 2-Chat for task-specific enhancements, employing techniques such as Instruction Fine-tuning, which leverages {prompt, response} pairs to improve adherence to textual instructions.
Due to the high resources required for full fine-tuning of all model layers, Parameter-Efficient Fine-Tuning (PEFT) methods like LoRA~\cite{hu_lora_2021} and its quantized version, QLoRA~\cite{dettmers_qlora_2023}, which selectively update parameters, have gained popularity. \textbf{Prompt Engineering (PE)} is another method to direct language models towards desired outputs by iteratively refining the input for a generative model, circumventing the need for extensive annotated data or altering model parameters.
\textbf{Few-shot prompting} includes examples in the prompt, giving the model additional context which aids in boosting its performance by guiding the model in generating outputs that mirror the patterns in the examples \cite{noauthor_llm_nodate}.
LLMs have found diverse applications in the field of cybersecurity. For example, ChatIDS~\cite{juttner_chatids_2023} utilizes an LLM to interpret and explain anonymized intrusion detection system alerts to non-experts, providing user-friendly explanations and demonstrating its understanding of cybersecurity concepts. Additionally, Fayyazi et al.\cite{fayyazi_uses_2023} explored using LLMs, such as GPT-3.5 and Bard, along with supervised training-based BaseLLMs, to classify cybersecurity descriptions into MITRE ATT\&CK tactics. The MITRE ATT\&CK framework~\cite{mitre_web}, a freely available knowledge base, categorizes adversary tactics and techniques to aid in understanding and defending against cyber threats.
Fayyazi et al.~\cite{fayyazi_uses_2023} highlighted challenges related to TTP description ambiguity and the importance of precise prompts.
While LLMs have recently been successfully applied to entity and relationship extraction in various domains \cite{li_instruction-tuned_nodate}\cite{wadhwa_revisiting_2023}\cite{zhang_aligning_2023}\cite{wan_gpt-re_2023}, their specific application in extracting structured information from CTI reports within the cybersecurity field has received less attention. Siracusano et al.~\cite{siracusano_time_2023} employed GPT-3.5 for CTI extraction. They implemented zero-shot prompting and in-context learning in two custom pipelines to develop a structured CTI extraction tool, streamlining the manual information extraction process. 

\textbf{Knowledge Graphs (KGs).} KGs are structured representations of real-world information in the form of triples $\boldsymbol{〈e_{head}, r, e_{tail}〉}$, where $\boldsymbol{e_{head}}$ denotes the head entity, $\boldsymbol{e_{tail}}$ the tail entity and $\boldsymbol{r}$ the relationship between the entities \cite{rossi_knowledge_2021}. Their ability to model structured, complex data in a machine-readable format makes KGs indispensable across diverse domains such as question-answering or information retrieval \cite{alam_looking_2023}. 
In CTI, KGs structurally represent complex relationships and integrate diverse data sources to enhance semantic querying, data analysis, and decision-making, thereby offering superior situational awareness and predictive insights.
The SEPSES Cybersecurity Knowledge Graph~\cite{ghidini_sepses_2019}, for example, combines comprehensive information on attack patterns and vulnerabilities from publically accessible sources into a dynamic KG.
Frameworks such as APTKG~\cite{sun_aptkg_2022} and AttacKG~\cite{li_attackg_2022} focus on the automated extraction and organization of threat intelligence entities and their relationships from CTI reports into KGs, aiming to improve the understanding of attack methods. THREATKG~\cite{gao_threatkg_2022} expands on this by automating the collection and integration of open-source threat intelligence into a KG, which is continually updated with new information. These studies highlight KGs' role in organizing CTI for advanced security analytics and forecasting. TINKER~\cite{rastogi_tinker_2023} and LADDER~\cite{alam_looking_2023} convert textual attack patterns into MITRE ATT\&CK-aligned structured data, further applying link prediction on the resultant CTI KGs.

\textbf{Link Prediction (LP).} LP addresses KG incompleteness by predicting missing facts \cite{rossi_knowledge_2021}. It identifies the missing entity in a relationship (head or tail prediction), completing a triplet either as the subject (head prediction) in the format $\boldsymbol{〈?, r, e_{tail}〉}$ or as the object (tail prediction) in $\boldsymbol{〈e_{head}, r, ?〉}$, effectively completing the given triplet in the KG.
KG embedding techniques like RotatE~\cite{rotate_2019}, ConvE~\cite{Jiang_convr_2019}, and TuckER~\cite{balazevic_tucker_2019} (used in tools like TINKER~\cite{rastogi_tinker_2023} and LADDER~\cite{alam_looking_2023}) have proven successful on benchmarks.
LP can be performed in a transductive setting, using only entities seen during training, or in an inductive setting, which accommodates unseen entities during testing through additional inference statements. \cite{ali2021improving}. While TuckER~\cite{balazevic_tucker_2019} is designed for transductive LP, newer methods like NodePiece~\cite{galkin2022nodepiece} are equipped to handle both settings. Link prediction in cybersecurity KGs enables proactive threat detection by forecasting potential future connections, like new malware attack patterns \cite{alam_looking_2023}. It aids vulnerability assessment by revealing possible attack paths, helping organizations strengthen their defenses. It also improves incident response and risk management by pinpointing and prioritizing potential threat vectors \cite{liu2022review}.

\section{Approach}
\subsection{Research Questions}

We address the following research questions (RQs) in this study through extensive experiments and analysis:

\begin{enumerate}[leftmargin=12pt]
\item \textbf{RQ1(Few-Shot):} How effective are LLMs at extracting CTI information in a few-shot setting? \textit{For this}, we investigate prompt engineering \& guidance frameworks \cite{noauthor_guidance-aiguidance_2024}.
\item \textbf{RQ2 (Fine-tuning):} How much does fine-tuning LLMs with labeled data improve CTI information extraction? \textit{For this}, we employ the parameter-efficient QLORA method.
\item \textbf{RQ3 (Knowledge Graph Quality):} What is the quality of triples generated from a large CTI corpus using LLMs, and how can we improve them? \textit{For this}, we identify shortcomings and propose error reduction methods.
\item \textbf{RQ4 (Link Prediction):} How does CTI-derived knowledge graph perform in link prediction? \textit{For this}, we consider both transductive and inductive settings.
\end{enumerate}

\subsection{Pretrained Models}

We selected several open-source pretrained LLMs based on their diverse capabilities. The Llama 2 series (7B, 7B chat, 13B, 13B chat, 70B chat) \cite{touvron_llama_2023}, Mistral 7B Instruct v2.0~\cite{jiang_mistral_2023}, and Zephyr-7B-\textbeta~\cite{tunstall_zephyr_2023} were included in the analysis (summary in Table~\ref{tab:model_summary}).
The Llama 2 series, introduced by Meta AI in February 2023, offers a range of pre-trained and fine-tuned LLMs with capacities ranging from 7 to 70 billion parameters. This variety enables a comprehensive analysis of techniques at different scales \cite{touvron_llama_2023}. Notably, the Llama 2 70B model outperforms MosaicML Pretrained Transformer (MPT) and Falcon \cite{touvron_llama_2023}. Llama 2-Chat models are fine-tuned with supervised fine-tuning and further trained with reinforcement learning with human feedback (RLHF) to align model behavior with human preferences and instruction following. They exhibit strong temporal knowledge organization abilities even with limited data, and its 70B version surpasses ChatGPT in answering factual questions \cite{touvron_llama_2023}.
Mistral 7B Instruct, fine-tuned on instruction datasets from the Hugging Face repository, outperforms smaller Llama 7B and 13B versions \cite{jiang_mistral_2023}. Additionally, Zephyr, based on Mistral 7B and fine-tuned with distilled Direct Preference Optimization (dDPO) to improve intent alignment, outperforms Llama 70B chat in MT-Bench tests \cite{tunstall_zephyr_2023}. 

\renewcommand{\arraystretch}{1.2}
\begin{table}[t]
    \centering
    \captionsetup{font=scriptsize} 
    \caption{Summary of Selected Models}
    \label{tab:model_summary}
    \tiny
\resizebox{\columnwidth}{!}{
    \begin{tabular}{ @{\hspace{2pt}}
        >{\raggedright\arraybackslash}p{0.8cm} >{\raggedright\arraybackslash}p{0.3cm} >{\raggedright\arraybackslash}p{1cm} >{\raggedright\arraybackslash}p{3.4cm}
        >{\raggedright\arraybackslash}p{0.5cm}  @{\hspace{6pt}}
    }
        \hline
        \textbf{Model} & \textbf{Size} & \textbf{Version} & \textbf{Architecture} & \textbf{Align} \\
        \hline
        Llama 2 
            & 7B  & Base & \multirow{4}{=}{Enhanced Llama 1 with doubled context length and expanded pre-training data} & -  \\
            \cline{2-3} \cline{5-5}
            & 7B & Chat & & RLHF \\
            \cline{2-3} \cline{5-5}
            & 13B & Base &   & - \\
            \cline{2-3} \cline{5-5}
            & 13B & Chat &  & RLHF \\
            \cline{2-5}
            & 70B & Chat & Enhanced Llama 1 with doubled context length, expanded pre-training data, and GQA  & RLHF \\
        \hline
        Mistral & 7B & Instruct v2.0 & GQA & - \\
        \hline
        Zephyr & 7B & \textbeta & Fine-tuned version of Mistral 7B v0.1 & dDPO \\
        \hline
    \end{tabular}
    }
\end{table}

\subsection{Dataset Generation}
\label{sec:dataset}

We utilized two datasets, introduced by Alam et al. in \cite{alam_looking_2023}, for our extraction experiments with LLMs: a manually annotated dataset for model training and evaluation and a large-scale dataset for knowledge graph (KG) generation and link prediction. The fine-tuning and evaluation dataset contains 120 curated CTI reports related to 36 Android malware families between the years 2015 and 2022.
The reports were manually annotated using the BRAT annotation tool, capturing cyber threat concepts such as \code{Malware}, \code{Malware Type}, \code{Application}, \code{Operating System}, \code{Organization}, \code{Person}, \code{Time}, \code{Threat Actor}, \code{Location}, \code{Indicator}, and \code{Attack Pattern}. These concepts are connected by ten relations: \code{isA}, \code{targets}, \code{uses}, \code{hasAuthor}, \code{hasAlias}, \code{indicates}, \code{discoveredIn}, \code{exploits}, \code{variantOf}, \code{has}. 

To accommodate the model's maximum content length of 4096 tokens and address the limited number of examples, we divided the text from annotated CTI reports into paragraphs. Various dataset sizes were tested, ranging from 400 to 1000 tokens per training example, resulting in different numbers of training examples. For example, for 400 tokens, the data was divided into 909 paragraphs. The annotated triples were matched automatically with the corresponding paragraphs via a script. However, not all paragraphs contained content that could be directly associated with the annotated triples. As a result, only 768 paragraphs successfully matched triples and were included in the final dataset. The dataset was split into training, validation, and test sets in an 80:16:4 ratio, yielding 574 training, 115 validation, and 29 test examples, each with up to 400 tokens per paragraph.

The best-performing LLMs, specifically the fine-tuned Llama 2 7B chat and the Llama 70B chat guidance model, were employed to generate a KG from the second, larger dataset comprising approximately 12,000 unstructured open-access CTI reports. These reports were divided into paragraphs, each limited to a maximum of 1,000 tokens, resulting in around 80,000 paragraphs. These paragraphs were then processed by the LLM to generate triples for the KG.

\subsection{Information Extraction Methods using LLM}

\subsubsection{Few-Shot Setting}
Two methods for extracting triples under a few-shot setting with LLMs are explored -- prompt engineering and the guidance framework.

\paragraph{\textbf{Prompt Engineering}}
\label{sec:pe}
To explore the effectiveness of prompt engineering, experiments were conducted with various chat variants of the selected models, including Llama 2 7B chat, 13B chat, and 70B chat, as well as Mistral 7B Instruct v2.0 and Zephyr-7B-\textbeta. 
Prompts were tailored to test the model's adaptability across varying instruction styles and complexity, covering scenarios from zero to few-shot inference with diverse formats and content. 
We followed the prompt engineering guidelines from OpenAI \cite{openai_prompt}\cite{openai_bestpractices}, which contain best practices on how to give clear and effective instructions and share strategies and tactics for getting better results from LLMs.
For example, including the ontology and instruction in the system prompt      
    \code{[INST] <<SYS>>\textbackslash nExtractct cybersecurity-related triples consisting of entities of the types Malware, Malware Types, Applications, Operating Systems, Organizations, Persons, Times, Threat Actors, Locations, and Attack Patterns and relationships between these entities of the types isA, targets, uses, hasAuthor, hasAlias, indicates, discoveredIn, exploits, variantOf, and has. Print the extracted triples in the format: [Entity1, Relation, Entity2]\textbackslash n<<SYS>>\textbackslash n\textbackslash nExtractct triples from the following text:``\{input\_txt\}'' [/INST]} 
compared to only including the relationships or no information about the ontology in the prompt, like 
    \code{[INST] What are the [subject, predicate, object]-triples in the following text? Text:``\{input\_txt\}'' [/INST]}.
Additionally, the number and format of the included examples, such as \code{Input text: ``A new version of the SpyNote Trojan is designed to trick Android users into thinking it’s a legitimate Netflix application. Once installed, the remote access Trojan (RAT) essentially hands control of the device over to the hacker, enabling them to copy files, view contacts, and eavesdrop on the victim, among other capabilities.``}

\code{Extracted triples: [SpyNote, isA, Trojan], [SpyNote, targets, Andriod], [SpyNote, uses, designed to trick Android users into thinking it’s a legitimate Netflix application], [SpyNote, isA, remote access Trojan], [SpyNote, isA, RAT], [SpyNote, uses, hands control of the device over to the hacker], [SpyNote, uses, enabling them to copy files], [SpyNote, uses, view contacts], [SpyNote, uses, eavesdrop on the victim]} varied between zero to four, separated by only line breaks, or the predefined prompt format, e.g., for Llama 2 \code{Input text: \{ex\_txt\} [/INST] Extracted triples: \{ex\_triples\} </s><s>[INST] Input text: \{text\} [/INST]}.

\paragraph{\textbf{Guidance}}
Guidance~\cite{noauthor_guidance-aiguidance_2024} is a Python library by Microsoft Research for controlling LLM output, using a special guidance language for applying constraints like regular expressions or Context-free grammar (CFGs). It constructs an Abstract Syntax Tree (AST) for program interaction, customizing LLM responses and stopping points. \cite{rechia_deep_2023}.
Additionally, the guidance framework boosts efficiency and accuracy by batching user-added text and omitting non-essential parts \cite{noauthor_guidance-aiguidance_2024}.
\par
In this research, the \textit{guidance} library is employed to enforce the desired output format for the Llama 2-Chat models. Its capability to ensure data structure compliance and seamless integration into data processing tasks makes it a valuable tool. 
Using a regular expression such as \code{\textbackslash\textbackslash n|</s>} as a stopping criterion makes the generation stop upon encountering either a line break or an end-of-sentence token generated by the LLM. However, guidance currently lacks support for negative or positive lookaheads and generates parsing errors in regular expressions, especially with special operator symbols like the closing bracket \code{]}, greatly limiting its potential applications. Consequently, in our experiments, only the stop regular expression is effectively utilized.
Instructions and examples can be incorporated as follows, once the model has been loaded and referenced as \code{llama2}: 
\begin{mycode}
def instruction(lm, ex_txt, ex_triple, input_txt):
    lm += f'''Extract triples from the following input text. Your answers need to be in the format [subject, predicate, object].
    -----------
    Input text: {example_txt}
    Extracted triples: {example_triple}
    -----------
    Input text: {input_txt}
    Extracted triples: '''
    return lm

output = llama2 
         + instruction(ex_txt, ex_triples, input_txt) 
         + gen(name="triples", temperature=0.6, regex=r'.+', 
               max_tokens=2048, stop_regex='\\n|</s>')
\end{mycode}

\subsubsection{Fine-tuning}
\label{method:ft}
While prompt engineering can yield significant improvements by optimizing prompts, fine-tuning a language model can enhance outcomes in specific scenarios and increase robustness. 
Therefore, a subset of the experiments focused on instruction fine-tuning the Llama 2 models (7B, 7B chat, 13B, and 13B chat) using a specific training dataset described in section \ref{sec:dataset}. 
QLoRA with 4-bit quantization was implemented to fine-tune the models in a resource-efficient manner.
The experiments encompassed several factors, including different prompts, dataset sizes (ranging from 357 paragraphs with 1000 tokens each to 768 paragraphs with 400 tokens each), training configurations (using linear, constant, or cosine learning schedulers with learning rates of 2e-4 or 2e-5), and LoRA parameters (with ranks and alphas of 8, 16, 32, or 64), allowing for a comprehensive exploration of the fine-tuning process.
The prompts experimented with various text separations from the instructions, employing line breaks, quotes, and instructional symbols such as \code{<txt></txt>} and \code{[TXT][/TXT]}, or using no separation at all.
Additionally, the complexity and length of the prompt were varied, ranging from  the inclusion of comprehensive ontology information, e.g., 
    \code{[INST] Extract cybersecurity-related triples from the following text. Present them in the structure [subject, predicate, object]. The following concepts should be used for subjects/objects: Malware, Malware Type, Application, Operating System, Organization, Person, Time, Threat Actor, Location, and Attack Pattern. The concepts should be connected through the following ten relationships as predicate: isA, targets, uses, hasAuthor, hasAlias, indicates, discoveredIn, exploits, variantOf, and has. The input text is: ``\{input\_txt\}'' [/INST]}
    to prompts with only a brief task instruction, like
    \code{[INST] Extract [subject, predicate, object]-triples from the following input text. Input text: ``\{input\_txt\}'' [/INST]}.
For fine-tuning the base models, the commonly used Alpaca prompt format \cite{alpaca} was employed in the form: 
\code{\#\#\# Instruction: Extract [subject, predicate, object]-triples from the following input text.\textbackslash n\#\#\# Input: \{input\_txt\}\textbackslash n\#\#\# Response: \{triples\}}

\subsubsection{Inference \& Decoding Strategies}
Transformer models predict next token by producing a probability distribution across all possible words, where decoding strategies crucially impact text coherence and repetition.
Greedy decoding risks repetition by choosing the likeliest next token, while beam-search considers multiple paths for higher-quality text \cite{hf_howtogenerate}.
Beam-search multinomial sampling blends the randomness of multinomial sampling with beam-search's strategy to optimize text generation \cite{hf_textgenerationstrategy}.
Adjusting decoding strategies and parameters like repetition penalty can improve text quality without changing the model's core settings.
The mentioned decoding strategies have been tested for prompt engineering and fine-tuning in the context of structuring CTI. The \textit{guidance} library currently does not support beam-search decoding or sampling strategies.

\subsection{Implementation Details}

To handle the computational demands of these large models, the experiments utilized NVIDIA A100 Tensor Core GPUs within a SPORC (Scheduled Processing On Research Computing) High-Performance Computing (HPC) cluster.

The models were loaded from Hugging Face~\cite{huggingface} using the \textit{transformers} library \cite{transformers}. For the quantization of smaller models, \textit{bitsandbytes} \cite{bitsandbytes} was utilized, whereas the Llama 2 70B chat model was used in its quantized form with \textit{autogptq}~\cite{autogptq}. The \textit{peft} \cite{peft}  and \textit{trl}~\cite{trl} library were employed for LoRA fine-tuning, and applied to all linear layers. Additionally, \textit{Accelerate} \cite{accelerate} was used to enhance inference speed during KG generation. For comparison, the maximum token length was set to 2048 tokens, which proved sufficient given the small input size of 1000 tokens. The temperature and repetition penalty for all models were set at 0.6 and 1.1, respectively. Decoding strategies and few-shot examples are detailed in Table \ref{tab:model_parameters}.

\begin{table}[htbp]
\centering
\captionsetup{font=scriptsize} 
\caption{Overview of LLM Inference Parameters: The context length is set at 2048 tokens for all models, the temperature is set to 0.6, and repetition penalty to 1.1.}
\label{tab:model_parameters}
\resizebox{\columnwidth}{!}{
\tiny
\begin{tabular}{@{\hspace{2pt}}lllllc@{\hspace{2pt}}}
\hline
\textbf{Technique} && \textbf{Model} && \textbf{Decoding Strategy} & \textbf{E.g. Number} \\
\hline
        PE 
        && Llama 2 models && Multinomial Sampling & 1-2 \\
        \cline{3-6}
        && Mistral 7B Instruct && Greedy & 1 \\
        \cline{3-6}
        && Zephyr-7B-\textbeta && Greedy & 2 \\
        \hline
        Guidance && Llama 2 models && Greedy & 1-2 \\
        \hline
        Fine-tuning && Llama 2 models && Beam-search & 0 \\
\hline
\end{tabular}
}
\end{table}

\subsection{Evaluation Metrics}

Traditional evaluation metrics, such as Accuracy and F{\scriptsize 1}-score, where only exact matches are compared, are inadequate for assessing the performance of generative models. Since these models generate new content based on learned patterns, they require more nuanced metrics that capture the quality and relevance of the generated text.
We adopt the ROUGE metrics \cite{lin_rouge_nodate} in our study to evaluate the quality of triples generated by LLMs. \textbf{ROUGE} (Recall-Oriented Understudy for Gisting Evaluation) \cite{lin_rouge_nodate} comprises a suite of metrics designed to compare generated text with reference text, making it suitable for applications where such comparisons are feasible, including translation, text summarization, and entity extraction. The effectiveness of ROUGE heavily depends on the quality of the reference text; poor-quality references can lead to misleading assessments of the generated content's quality. It is case-insensitive and produces values ranging from 0 to 1. ROUGE-N evaluates the overlap of n-grams (unigrams, bigrams, and trigrams) between a text and its reference, using metrics like precision, recall, and F{\scriptsize 1}-score. Recall measures how many relevant words from the reference text are captured by the machine-generated text, emphasizing the inclusion of all pertinent information. 
The recall of n-gram overlaps between a candidate text and a set of reference texts is calculated as follows: 
\begin{equation}
    \scriptsize
    \text{ROUGE-N}_{recall} = \frac{\sum_{S \in \{\text{Reference}\}} \sum_{gram_n \in S} \text{Count}_{\text{match}}(gram_n)}{\sum_{S \in \{\text{Reference}\}} \sum_{gram_n \in S} \text{Count}(gram_n)}
\end{equation}

Precision assesses the proportion of words in the machine-generated text that accurately match those in the reference text, focusing on the relevance and correctness of the information provided. It is calculated with the following formula:
\begin{equation}
    \scriptsize
    \text{ROUGE-N}_{precision} = \frac{\sum_{S \in \{\text{Reference}\}} \sum_{gram_n \in S} \text{Count}_{\text{match}}(gram_n)}{\sum_{S \in \{\text{Canidate}\}} \sum_{gram_n \in S} \text{Count}(gram_n)}
\end{equation}

The F{\scriptsize 1}-score balances recall and precision, offering a single metric to assess overall performance:

\begin{equation}
    \scriptsize
    \text{F}_{1}\text{-score} = \frac{2 * P * R}{P + R}
\end{equation}

Considering the reference hand-annotated triple \texttt{\small[\underline{Adwind}, \underline{targets}, retail and \underline{petroleum} \underline{industry}]} and the candidate \texttt{\small[\underline{Adwind}, \underline{targets}, \underline{petroleum} production \underline{industry}]} extracted with the LLM, the ROUGE-1 recall would be $4/6 = 0.667$, representing four unigram matches out of six possible unigrams in the reference. Subsequently, the precision would be calculated as $4/5 = 0.8$, and F{\scriptsize 1}-score $2*0.667*0.8 / 1.467 = 0.727$, which is reported in this paper.
It is important to note that for ROUGE-N the order in which different n-grams appear in the text does not influence the score. However, the words within an individual n-gram must be consecutive. For example, in ROUGE-2, the matches would only include \texttt{\small[\underline{Adwind, targets,} retail and petroleum industry]} from the reference and \texttt{\small[\underline{Adwind, targets,} petroleum production industry]} from the candidate.
In contrast, ROUGE-L assesses the longest common subsequence between texts, capturing similarity in content without requiring consecutive word matches but in-sequence matches. This allows ROUGE-L to reward summaries with similar content even if their sentence structures differ. Thus, in the given example, the reference and candidate would be matched as \texttt{\small[\underline{Adwind, targets,} retail and \underline{petroleum industry}]} and \texttt{\small[\underline{Adwind, targets, petroleum} production \underline{industry}]}, respectively.

To evaluate the model's performance on the test set, triples were extracted from the LLM's output using regular expressions and sorted alphabetically. The hand-annotated triples from the dataset, also sorted alphabetically, served as the reference text for comparison.
For the initial assessment, a manually evaluated example paragraph was used. Based on the desired outcome, the prompt and settings that produced the closest alignment with the desired output were applied to generate outputs for the remaining 29 examples in the test set. The generated results were evaluated using ROUGE metrics~\cite{lin_rouge_nodate} and human evaluation.

\section{Experiment and Results}
\subsection{Few-Shot Performance (RQ1)}

\textbf{Prompt engineering} significantly affects the output of LLMs, with elements such as prompt structure, spacing, line breaks, and punctuation playing a crucial role. 
The impact of the prompt design was evident when comparing outputs from different sizes of the Llama 2 model; a prompt that was highly effective with the 7B model yielded less satisfactory results with the 13B variant. For example, considering the one-shot prompt
     \code{[INST] <<SYS>>\textbackslash nExtractct [subject, predicate, object]-triples from the input text with the following predicates: isA, targets, uses, hasAuthor, hasAlias, indicates, discoveredIn, exploits, variantOf, and has.\textbackslash n<<SYS>>\textbackslash n\textbackslash nInputut text: \{ex\_txt\} [/INST] Extracted triples: \{ex\_triples\} </s><s>[INST] Input text: \{input\_txt\} [/INST] Extracted triples: }.
This prompt yielded the desired output as illustrated in Table \ref{tab:decoding}, where prompt engineering combined with greedy decoding produced the expected results for the 7B model. However, applying the same configuration to the 13B model resulted in merely listing entities from the text, e.g., \code{The Adwind RAT, petroleum industry, US, new campaign, multi-layer obfuscation, evading detection,...}.
Consequently, this makes it a time-consuming process of trial and error.
While adding a single example often enhanced the output, incorporating multiple examples, as the one mentioned in section \ref{sec:pe}, did not necessarily lead to better outcomes and, in some instances, even deteriorated the model's performance. For instance, the Llama 2 7B chat model produced continuous text with over two examples, whereas other models struggled with inaccurately inventing relationships like \code{hostedOn} or \code{usedFor} from text verbs.
Llama 2-Chat models presented notable challenges in generating outputs in a format suitable for further processing, in contrast to Mistral and Zephyr models. Llama models exhibited less consistency in output formatting, whereas Mistral, in particular, was able to produce the desired format reliably, even in zero-shot inference scenarios. 
The prompt \code{[INST] <<SYS>>\textbackslash nYouou are a helpful cyber-security assistant. Your task is to extract [entity,relationships,entity]-triples from the input text. Use only the relationships: `isA', `targets', `uses', `hasAuthor', `hasAlias', `indicates', `discoveredIn', `exploits', `variantOf', and `has'. Your answers need to be in the format [entity,relationship,entity]. Reply only with the extracted triples.\textbackslash n<</SYS>>\textbackslash n\textbackslash nWhatat are the triples in the following input text: \{input\_txt\} [/INST]} with Mistral greedy decoding already yielded triples in the format \code{[Adwind RAT,isA,Remote Access Trojan (RAT)],
 [Adwind RAT,targets,petroleum industry],
 [Adwind RAT,targets,US]}. 

To ensure the model maintained the desired relationship types in the triples, merely providing examples was insufficient; incorporating the ontology directly into the prompt proved to be crucial. 
The most effective results across all models were achieved by combining the ontology with one or two examples in the prompt, as illustrated in the prompt described in the first paragraph of this section.

Our comparative analysis of model performance with prompt engineering is summarized in Table \ref{tab:pe-ROUGE}, which presents the ROUGE metric scores in the context of n-gram overlap for n=1,2,3,6 (ROUGE-N), as well as the longest common subsequence (ROUGE-L) for different prompts and models applied to the extraction of triples from the test data. It lists the results of the best-performing prompts for the models following the triple extracting from the LLM's output. The ROUGE scores support the findings that Mistral and Zephyr outperform Llama 2 models when only prompt engineering is applied. 
The frequent failure of the Llama models to generate the desired triples led to a score of 0 for those specific test instances, which significantly reduced the overall ROUGE score. Noteworthy is the performance of the Llama 2 70B chat model, which, when prompted using a one-shot technique, achieved results that were the most comparable. However, its lower score can be attributed to its inability to fully comply with the specified ontology, particularly concerning the relationship types outlined in the prompt. This led to the generation of triples that diverged from the manually annotated reference, a discrepancy that became more pronounced in longer texts.

\begin{table}[htbp]
\centering
\captionsetup{font=scriptsize}
    \caption{Comparison of PROMPT ENGINEERING Model Performance on ROUGE Metrics: Evaluating Various Prompts and Models for Triple Extraction from Test Data. The highest value in each metric is in bold.}
    \label{tab:pe-ROUGE}
    \resizebox{\columnwidth}{!}{
\begin{tabular}{@{}llcccccc@{}}
\toprule
\textbf{Model} & \textbf{Prompt} & \textbf{ROUGE-1} & \textbf{ROUGE-2} & \textbf{ROUGE-3} & \textbf{ROUGE-6} & \textbf{ROUGE-L} \\ \midrule
        Llama 2 7B chat& few-shot & 0.3328 & 0.2012 & 0.1199 & 0.0357 & 0.2407 \\
        Llama 2 13B chat& one-shot & 0.3825 & 0.2336 & 0.1452 & 0.0450 & 0.2910 \\
        Llama 2 13B chat& few-shot & 0.4205 & 0.2505 & 0.1521 & 0.0521 & 0.3313 \\
        Llama 2 70B chat& one-shot & 0.3930 & 0.2414 & 0.1561 & 0.0674 & 0.3040 \\
        Mistral 7B Instruct& one-shot & 0.4775 & \textbf{0.3170} & \textbf{0.2083} & \textbf{0.0833} & \textbf{0.3388} \\
        Zephyr-7B-\textbeta & few-shot & \textbf{0.4332} & 0.2856 & 0.1968 & 0.0796 & 0.3209 \\ 
\bottomrule
\end{tabular}
}
\end{table}

\textbf{Guidance} demonstrated remarkable efficiency, delivering impressive outcomes and proving easy to implement. It surpassed prompt engineering in achieving higher ROUGE scores, as illustrated in Table \ref{tab:guid-ROUGE}, which displays the scores for various models using \textit{guidance}. For the smaller Llama 2-Chat models, optimal performance was achieved using the prompt 
    \code{Extract triples from the following input text. Your answers need to be in the format [entity1, relation, entity2].} 
with three examples, whereas the 70B chat model showed its best performance with a more extensive prompt that included the ontology, i.e.,
    \code{Extract [entity, relationships, entity]-triples from the following input text. Only use the relationships: `isA', `targets', `uses', `hasAuthor', `hasAlias', `indicates', `discoveredIn', `exploits', `variantOf', and `has'.}
and just one example.
Although the 13B chat model attained the highest ROUGE scores, manual human evaluation of the test data's generated triples exposed its difficulty with complex syntactical structures such as negations, resulting in the majority of triples being incorrect. For instance, from the text \code{First Twitter‑controlled Android botnet discovered Detected by ESET as Android/Twitoor, this malware is unique because of its resilience mechanism. Instead of being controlled by a traditional command-and-control server, it receives instructions via tweets.} the model incorrectly extracts triples such as \code{[Twitoor, isA, command-and-control server]} and \code{[Twitoor, isA, traditional command-and-control server]}, failing to correctly interpret the word \code{instead}.

The same issue was observed with the 7B model, positioning the 70B model as the top-performing guidance model according to human analysis.

\begin{table}[htbp]
\centering
\captionsetup{font=scriptsize}
\caption{Comparison of GUIDANCE Model Performance on ROUGE Metrics: Evaluating Various Prompts and Models for Triple Extraction from Test Data. The highest value in each metric is in bold.}
\label{tab:guid-ROUGE}
\resizebox{\columnwidth}{!}{
\begin{tabular}{@{}llcccccc@{}}
\toprule
\textbf{Model} & \textbf{Prompt} & \textbf{ROUGE-1} & \textbf{ROUGE-2} & \textbf{ROUGE-3} & \textbf{ROUGE-6} & \textbf{ROUGE-L} \\ \midrule
Llama 2 7B chat & few-shot         & 0.4724  & 0.3025  & 0.1930  & 0.0635  & 0.3437  \\
Llama 2 13B chat & one-shot        & \textbf{0.5263}  & \textbf{0.3827}  & \textbf{0.2898}  & \textbf{0.1516}  & \textbf{0.4355}  \\
Llama 2 70B chat & one-shot        & 0.5147  & 0.3730  & 0.2791  & 0.1425  & 0.3999  \\
Llama 2 70B chat & few-shot        & 0.5126  & 0.3527  & 0.2493  & 0.1137  & 0.3905  \\ \bottomrule
\end{tabular}
}
\end{table}

\begin{highlighted}{}
\footnotesize
\textbf{Finding 1: }Guidance performs better than simple prompt engineering in a few-shot setup for CTI triple extraction and requires less effort to achieve accurate outputs. 
\end{highlighted}

\subsection{Fine-tuning Performance (RQ2)}

\textbf{Fine-tuning} with fewer than the 718 examples in the dataset proved insufficient. The most effective approach utilized a constant learning scheduler with a learning rate of 2e-4, and LoRA parameters set with an alpha and rank of 16 for all linear layers. 
Significant variations in output were observed when different symbols were used to separate the input text from the prompt.
Introducing line breaks into the input text for fine-tuning the model led to poor results, characterized by outputs that were either mere numbers or repetitive sections of the prompt. This underscores the critical importance of strict adherence to formatting, particularly regarding line breaks and prompt structure, to attain the intended output.
Moreover, longer instructions tended to confuse the language model. 
For instance, the prompt 
    \code{[INST] Extract [subject, predicate, object]-triples from the following input text. Input text: ``\{input\_txt\}'' [/INST]} 
significantly outperformed more detailed prompts that included the ontology, as discussed in section \ref{method:ft}. The latter failed to extract any triples, merely generating repetitive text regardless of the decoding strategy employed.

As illustrated in Table \ref{tab:decoding}, which presents a comparison of the different outputs generated by applying various decoding strategies in the context of Llama 2 7B chat prompt engineering versus fine-tuned model, the impact on the fine-tuned model is more pronounced. 
While greedy decoding restricted the output to only one word and multinomial sampling to only two triples with the fine-tuned model, beam-search multinomial sampling resulted in greater variance but highly repetitive output towards the end, even with a repetition penalty of 1.1. Generally, beam-search outperformed other strategies for fine-tuning, while prompt engineering achieved the desired output with greedy decoding and multinomial sampling, as evident in the table with the most relevant triples generated for those strategies. To avoid repetitive output, setting a repetition penalty of 1.1 for beam-search was necessary.

\begin{table*}[!ht]
    \centering
    \captionsetup{font=scriptsize} 
    \caption{Comparison of Decoding Strategies: Llama 2 7B chat Prompt Engineering compared with Fine-tuned Model Across Various Decoding Techniques}
    \label{tab:decoding}
    \tiny
    \begin{tabular}{
    >{\raggedright\arraybackslash}p{1cm} >{\raggedright\arraybackslash}p{7.45cm} >{\raggedright\arraybackslash}p{7.45cm}
    }
    \hline
    \textbf{Decoding Strategy} & \textbf{Prompt Engineering} & \textbf{Fine-tuned Model} \\
    \hline
    \textbf{Greedy} & \texttt{[Adwind RAT, isA, RAT], [Adwind RAT, targets, petroleum industry in the US], [Adwind RAT, uses, multi-layer obfuscation to evade detection], [Adwind RAT, isA, malware], [Adwind RAT, has, persistence through registry modifications], [Adwind RAT, performs, process injection to stay under the radar], [Adwind RAT, terminates, security services (e.g., firewall, AV)], [Adwind RAT, steals, sensitive data]} & \texttt{nobody} \\
    \hline
    \textbf{Beam-search} 
    &  \texttt{1. Adwind RAT, isA, malware \linebreak 
    2. Adwind RAT, targets, petroleum industry in the US \linebreak 
    3. Adwind RAT, uses, multi-layer obfuscation to evade detection \linebreak 
    4. Adwind RAT, is hosted on a serving domain \linebreak 
    ... \linebreak
    10. Adwind RAT, steals sensitive data.}
    &  \texttt{['Adwind', 'targets', 'petroleum industry'], ['Adwind', 'targets', 'US'], ['Adwind', 'targets', 'retail'], ['Adwind', 'targets', 'hospitality'], ['Adwind', 'uses', 'achieves persistence through registry modications'], ['Adwind', 'uses', 'performs process injection to stay under the radar'], ['Adwind', 'uses', 'terminates security services (e.g., rewall, AV)'], ['Adwind', 'uses', 'steals sensitive data']} \\
    \hline
    \textbf{Multinomial Sampling} 
    & \texttt{[Adwind RAT, isA, malware]}\linebreak
    \texttt{[Adwind RAT, targets, petroleum industry in the US]}\linebreak
    \texttt{[Adwind RAT, uses, multi-layer obfuscation to evade detection]}\linebreak
    \texttt{...} \linebreak
    \texttt{[Adwind RAT, uses, steals sensitive data]}
    & \texttt{['Adwind', 'targets', 'United States'], ['Adwind', 'isA', 'RAT']} \\
    \hline
    \textbf{Beam-search Multinomial Sampling} & 
    \texttt{1. Adwind RAT, isA, malware \linebreak 
    2. Adwind RAT, targets, petroleum industry in the US \linebreak 
    3. Adwind RAT, uses, multi-layer obfuscation to evade detection \linebreak 
    4. Adwind RAT, is hosted on a serving domain \linebreak 
    ... \linebreak
    10. Adwind RAT, steals sensitive data.}
    & \texttt{['Adwind', 'targets', 'petroleum'], ['Adwind', 'targets', 'US'], ['Adwind', 'targets', 'retail'], ['Adwind', 'targets', 'hospitality'], ['Adwind', 'uses', 'achieves persistence through registry modications'], ['Adwind', 'uses', 'performs process injection to stay under the radar'], ['Adwind', 'uses', 'steals sensitive data'], ['Adwind', 'targets', 'US'], ['Adwind', 'targets', 'petroleum'], ['Adwind', 'targets', 'retail'], ['Adwind', 'targets', 'hospitality'], ...}repeatedly\\
    \hline
\end{tabular}
\end{table*}

Fine-tuning the base models resulted in repetitive output that did not align with the instructions or the desired outcome, suggesting that the dataset size was insufficient for instruction-based tuning of the base Llama 2 models. Notably, loading the trained LoRA weights from the base models into the chat version for inference yielded good results, comparable to fine-tuning the chat versions directly. This suggests that chat-optimized models, being more adept at handling instructional tasks, can effectively capture essential patterns or knowledge relevant to this application through the LoRA weights, even if they were trained using a different prompt format.
\par
The ROUGE scores for the Llama 2 7B and 13B chat versions, differentiated by the use of either \code{<txt>} and \code{</txt>} markers or quotes to separate the input text from the prompt, alongside the fine-tuned base models employing the alpaca prompt format, are presented in Table \ref{tab:ft-ROUGE}. 
The fine-tuned chat models demonstrate notable consistency across the metrics, achieving higher scores than any other techniques could attain, highlighting the effectiveness of fine-tuning in improving the quality of model outputs.
Manual analysis of the generated triples from the test data supports the ROUGE scores. The fine-tuned Llama 2 7B chat model, utilizing input text marked by \code{<txt>} and \code{</txt>} , along with the fine-tuned Llama 2 7B and 13B models loaded onto their respective chat versions, exhibited the best performance. However, the latter models each failed to produce a response for different test samples.

\begin{table}[htbp]
\centering
\captionsetup{font=scriptsize}
\caption{Comparison of FINE-TUNED Model Performance on ROUGE Metrics: Evaluating Various Fine-tuning Prompt Formats and Models for Triple Extraction from Test Data. The highest value in each metric is in bold, with the second and third highest values underlined.}
    \label{tab:ft-ROUGE}
\resizebox{\columnwidth}{!}{
\begin{tabular}{@{}llccccc@{}}
\toprule
\textbf{Model} & \textbf{Prompt} & \textbf{ROUGE-1} & \textbf{ROUGE-2} & \textbf{ROUGE-3} & \textbf{ROUGE-6} & \textbf{ROUGE-L} \\ \midrule
        Llama 2 7B chat & txt & \textbf{0.6264} & \textbf{0.5419} & \textbf{0.4866} & \textbf{0.3404} & \textbf{0.5667} \\
        Llama 2 7B chat & quotes & 0.3976 & 0.3299 & 0.2749 & 0.1696 & 0.3508 \\
        Llama 2 7B on chat & alpaca & \underline{0.5965} & 0.5060 & \underline{0.4424} & 0.2842 & \underline{0.5469} \\
        Llama 2 13B chat & txt & 0.5786 & \underline{0.5107} & \underline{0.4522} & \underline{0.3162} & \underline{0.5353} \\
        Llama 2 13B chat & quotes & 0.5130 & 0.4443 & 0.3898 & 0.2761 & 0.4678 \\
        Llama 2 13B on chat & alpaca & \underline{0.6095} & \underline{0.5103} & 0.4389 & \underline{0.2868} & 0.5345 \\
\bottomrule
\end{tabular}
}
\end{table}

Given that the fine-tuned 7B chat model demonstrated superior performance in both the ROUGE metrics and manual analysis, it was selected for the subsequent phase: generating a KG from CTI data.

\begin{highlighted}{}
\footnotesize
\textbf{Finding 2: }Fine-tuning improves over few-shot prompting as evidenced by improved ROUGE score. Shorter prompts outperform longer ones and the separation format between instruction and input text significantly affects performance.
\end{highlighted}

\subsection{Knowledge Graph \& Enhancement (RQ3)}

Despite the model's optimal performance on the test data, as evidenced by both ROUGE scores and manual human evaluations, its application to a large-scale dataset for KG generation produced very noisy triples. 
In the initial approach, which involved generating triples from the text without naming entity types, the fine-tuned 7B model extracted 188,547 triples from nearly 12,000 documents. Upon first manual inspection, it became evident that the model generated numerous nonsensical triples. It not only extracted entities that might be correct but did not correspond to a given entity type, such as \code{[+44113320****, indicates, Phone Numbers]}, but also identified non-named entities, like \code{[124, indicates, Android]}, and produced factually incorrect triples, for instance 
\code{[/facebook, targets, Facebook]}. A qualitative assessment of 100 randomly sampled triples revealed  that only 39 of them were correct. Post-processing, for example, enforcing dates as the only valid subject for the \code{discoveredIn} relationship, or excluding numbers as subjects, was minimally effective and did not yield a usable KG.
This scenario illustrates that while LLMs may perform well on specific, curated test datasets, they struggle with large-scale, primarily unprocessed data. The limitations observed could be attributed partly to the relatively small size of the LLM, having only 7 billion parameters, and possibly to insufficient training data.

To facilitate data analysis of the generated triples and perform post-processing steps in relation to the ontology, additional experiments were conducted not only to extract the triples but also to identify the entity types within them. For instance, triples were generated in formats such as \code{[Adwind[Malware], targets, US[Location]]} and \code{[Adwind[Malware], isA, Trojan[MalwareType]]}

Table \ref{tab:ROUGE-type} summarizes the ROUGE scores for various techniques and models that showed promise in earlier stages by producing triples with entity types for the test data.
The 70B chat model failed to generate the triples in a format suitable for automatic processing in 19 of the 29 test examples, resulting in a low overall ROUGE score. Mistral and Zephyr achieved significantly better scores; however, manual analysis revealed that, although they extracted numerous triples, they frequently introduced incorrect relationships, such as \code{hasCapability} or \code{hasFeature}, and incorrect entity types, such as \code{Command} or \code{Attack}. Moreover, there were numerous errors in the entity types for the extracted triples, such as categorizing \code{APK} as \code{Indicator} and \code{banking apps} as \code{AttackPattern}.

The guidance models attained even higher ROUGE scores, but human evaluation of the generated indicated that they struggled with accurately extracting and classifying indicators, often misclassifying them as \code{AttackPattern}. Furthermore, when the extracted entity was correctly identified as \code{Indicator}, the relationship often did not adhere to the ontology, for example \code{[FakeSpy[Malware], isA, ANDROIDOS\_LOADGFISH.HRX[Indicator]]}. The gudiance models  also struggled with adhering strictly to predefined relationships.

The fine-tuned model was most accurate in adhering to the ontology, especially with \code{AttackPattern} and \code{Indicator} extraction and triple generation. However, it sometimes struggled to extract triples at all, merely repeating the input text, which resulted in lower ROUGE scores. Combining the fine-tuned model with the guidance framework during inference proved not to offer significant advantages.

\begin{table}[htbp]
\centering
\captionsetup{font=scriptsize}    
\caption{Comparison of Model Performance on ROUGE Metrics: Evaluating Various Techniques and Models for Extracting Triples including the ENTITY TYPES from the Test Data. The highest value in each metric is in bold, with the second and third highest values underlined.}
    \label{tab:ROUGE-type}
\resizebox{\columnwidth}{!}{
\begin{tabular}{@{}llcccccc@{}}
\toprule
\textbf{Technique} & \textbf{Model} & \textbf{ROUGE-1} & \textbf{ROUGE-2} & \textbf{ROUGE-3} & \textbf{ROUGE-6} & \textbf{ROUGE-L} \\ \midrule
        PE one-shot & Llama 70B chat & 0.1962 & 0.1433 & 0.1094 & 0.0534 & 0.1617 \\
        PE few-shot & Mistral 7B Instruct & 0.5094 & 0.3593 & 0.2639 & 0.1240 & 0.3874 \\
        PE few-shot & Zephyr-7B-\textbeta & 0.4986 & 0.3740 & 0.2873 & 0.1544 & 0.3824 \\
        
        Guidance few-shot & Llama 7B chat & \underline{0.5292} & \underline{0.4101} & \underline{0.3205} & \underline{0.1767} & \underline{0.4090} \\
        Guidance one-shot & Llama 13B chat & 0.5044 & 0.3744 & 0.2856 & 0.1502 & \underline{0.4207} \\
        Guidance one-shot & Llama 70B chat & \textbf{0.5565} & \textbf{0.4360} & \underline{0.3479} & \underline{0.2076} & \textbf{0.4372 }\\
        Guidance few-shot  & Llama 70B chat & \underline{0.5414} & \underline{0.4134} & 0.3193 & 0.1605 & 0.4018 \\
        
        FT txt & Llama 7B chat & 0.4351 & 0.3973 & \textbf{0.3623} & \textbf{0.2648} & 0.3857 \\
        FT txt + Guidance & Llama 7B chat & 0.5215 & 0.4031 & 0.3153 & 0.1752 & 0.4041 \\
\bottomrule
\end{tabular}
}
\end{table}

Based on these results, both the fine-tuned Llama 2 7B chat model and the 70B guidance model equipped with entity type extraction were chosen for generating the knowledge graph from the 12,000 documents.
Consistent with the performance observed on the test data, the fine-tuned extracted fewer triples, yielding a total of 77,307. The qualitative assessment revealed that 55 out of 100 randomly sampled triples were correct, indicating significant improvement compared to fine-tuning and extraction processes that did not utilize entity types. Post-processing efforts such as enforcing the specified ontology, excluding \code{Time} entities lacking a date, omitting triples where the \code{Malware} entity type was mentioned fewer than five times across all documents, and removing non-named entities from the \code{Malware} and \code{ThreatActor} categories, effectively reduced the number to 41,525 triples. Although the data still contained some noise, the qualitative assessment showed improvement with 77 out of 100 triples being correct. These refined triples were then used for link prediction.
The guidance model extracted nearly 460,000 triples from the documents, but the qualitative assessment revealed that the majority were incorrect, with only 15 out of 100 random samples being correct. After post-processing, only 17,050 entities remained, demonstrating higher quality, as evidenced by 66 correct samples in the qualitative assessment. This significant reduction in triples was due to the model incorrectly connecting entity types with certain relationships and extracting triples with newly invented entity types or relationships. Given that the fine-tuned model yielded better-quality triples, the triples from the guidance model were not utilized for further link prediction tasks.

\begin{highlighted}{}
\footnotesize
\textbf{Finding 3: }Applying models that perform well on a small-scale test set to a large-scale dataset can yield poor results with a lot of noise. To adhere to an ontology for triple extraction, including the entity types in the few-shot examples or fine-tuning data is necessary and produces output better suitable for post-processing.
\end{highlighted}

\subsection{Link Prediction (RQ4)}

For evaluating the efficacy of the prediction task, especially within knowledge graph forecasting, a comparison with triples generated using LADDER, as presented by Alam et al. \cite{alam_looking_2023}, based on the same corpus of 12,000 documents was conducted. Traditional metrics such as Mean Reciprocal Rank (MRR) and Hits@n were employed. These metrics derive from the rankings that the link prediction model assigns to all accurate test triples. MRR represents the mean of the inverse rankings of all correct test triples, while Hits@n indicates the proportion of instances where a true triple appears within the top n positions of the rankings. Higher values in these metrics indicate superior performance.

The evaluation utilized the TuckER \cite{balazevic_tucker_2019} model within a transductive setting across two distinct test sets. \textit{TestSet1} comprises hand-annotated triples that are not included in the training triples and have not been mapped to MITRE ATT\&CK techniques. Given TuckER's reliance on pre-identified entities, an additional test set containing only entities that were extracted by both LADDER and the fine-tuned 7B chat model was constructed as \textit{TestSet2}. TuckER was implemented using the \textit{pykeen} library \cite{ali2021pykeen}, adopting Alam et al.'s \cite{alam_looking_2023} parameters. The setup included 50 embedding dimensions, a batch size of 64, and an initial learning rate of 0.001, while the training was reduced to 500 epochs.

The results, as presented in Table \ref{tab:combined}, demonstrate that for \textit{TestSet1}, the TuckER model, when trained on triples extracted through the fine-tuned 7B chat model, achieved significantly higher scores, with a notable increase of 9.26\% in Hits@30 compared to its counterpart. For \textit{TestSet2}, the performance gap narrows, with the fine-tuned model exhibiting only marginal improvements in Hits@30 and MRR metrics. This suggests that the triples extracted using our methodology may be slightly more refined, thereby enhancing the model's ability to predict links between known entities more effectively. However, when comparing our results to those reported by Alam et al. \cite{alam_looking_2023}, it is evident that our overall scores are significantly lower. This discrepancy can be attributed to the absence of a mapping for MITRE ATT\&CK techniques, which leads to a reduction in the number of unique entities, consequently simplifying both the training and prediction processes. 

\begin{table}[!ht]
\centering
\captionsetup{font=scriptsize}
    \caption{Link Prediction Results with TuckER \& NodePiece for LADDER and Llama 2 7B fine-tuned chat model, including for Tail, Head, and Both results. H@ is Hits@.}
    \label{tab:combined}
\resizebox{1\columnwidth}{!}{
    \begin{tabular}{lcccccccccccc}
        & \multicolumn{4}{c}{\textbf{Head}} & \multicolumn{4}{c}{\textbf{Tail}} & \multicolumn{4}{c}{\textbf{Both}} \\
        \cmidrule(lr){2-5} \cmidrule(lr){6-9} \cmidrule(lr){10-13}
         & H@3 & H@10 & H@30 & MRR & H@3 & H@10 & H@30 & MRR & H@3 & H@10 & H@30 & MRR \\
        \midrule
        \multicolumn{13}{c}{Transductive Link Prediction TuckER TestSet-1} \\
        \midrule
        LADDER & .0194 & .0291 & .0922 & .0282 & .0269 & .0791 & .1279 & .0304 & .0364 & .1019 & .1869 & .0442 \\
        FT 7B & .0388 & .0631 & .1505 & .0390 & .0370 & .1077 & .2205 & .0451 &  .0437 & .1117 & .2379 & .0563 \\
        \midrule
        \multicolumn{13}{c}{Transductive Link Prediction TuckER TestSet-2} \\
        \midrule
        LADDER & .0135 & .0202 & .0303 & .0132 & .0534 & .1748 & .2816 & .0601 &  .0202 & .0497 & .0791 & .0218 \\
        FT 7B  & .0522 & .0875 & .1481 & .0530 & .0485 & .1602 & .3252 & .0735 & .0446 & .0976 & .1843 & .0491 \\
        \midrule
        \multicolumn{13}{c}{Inductive Link Prediction NodePiece TestSet-3} \\
        \midrule
        LADDER & .8545 & 1 & 1 & .6397 & .2909 & .6545 & 1 & .2442 & .5727 & .8273 & 1 & .4419 \\
        FT 7B  & .9273 & 1 & 1 & .7594 & .2909 & .6545 & .9636 & .2447 & .6091 & .8273 & .9818 & .5020 \\
        \bottomrule
    \end{tabular}
    }
\end{table}

\begin{table*}[!ht]
    \centering
    \captionsetup{font=scriptsize} 
    \caption{Case Study: A Comparison of the Llama 2 7B Fine-tuned Model and the Llama 2 70B Guidance Model in Extracting Triples from Two Example Paragraphs of Unstructured CTI Reports.}
    \label{tab:casestudy}
    \tiny
    \begin{tabular}{
    >{\raggedright\arraybackslash}p{6.2cm} >{\raggedright\arraybackslash}p{2.5cm} >{\raggedright\arraybackslash}p{7.2cm}
    }
    \hline
    \textbf{Paragraph} & \textbf{7B chat Fine-tuned} & \textbf{70B chat Guidance} \\
    \hline
     Leak of Alleged Pegasus Target List Restokes Spyware Debate Pegasus Spyware Critics See Shift to Mass Surveillance; Vendor NSO Group Disagrees Amnesty International on Sunday warned that Pegasus software was able to exploit Apple's latest iPhone 12 running the current 14.6 operating system. (Photo: Apple) Allegations that commercially available spyware is being abused by countries to spy on dissidents, journalists, political rivals and business leaders are again in the spotlight. See Also: Live Webinar | Improve Cloud Threat Detection and Response using the MITRE ATT\&CK Framework The leaking of an alleged target list tied to users of Israel-based NSO Group's Pegasus spyware - built to infect even the latest, fully patched Apple and Android devices - has led to questions over the scale of such operations, if the use of Pegasus gets sufficiently policed and whether the sale of spyware to certain countries should be blocked.
     & \texttt{[Pegasus[Malware], hasAuthor, Israel-based NSO Group[ThreatActor]], [Pegasus[Malware], targets, Apple[Organization]], [Pegasus[Malware], targets, iPhone 12[Hardware]], [Pegasus[Malware], targets, Android[OS]]}
     & \texttt{[Pegasus[MalwareType], isA, Spyware[MalwareType]], [Pegasus[MalwareType], targets, dissidents[Person]], [Pegasus[MalwareType], targets, journalists[Person]], [Pegasus[MalwareType], targets, political rivals[Person]], [Pegasus[MalwareType], targets, business leaders[Person]], [Pegasus[MalwareType], uses, exploit Apple's latest iPhone 12 running the current 14.6 operating system[AttackPattern]], [NSO Group[Organization], hasAuthor, Pegasus spyware[MalwareType]], [NSO Group[Organization], uses, built to infect even the latest, fully patched Apple and Android devices[AttackPattern]], [MITRE ATT\&CK Framework[Indicator], isA, Live Webinar[Event]], [MITRE ATT\&CK Framework[Indicator], isA, Improve Cloud Threat Detection and Response[Event]]} \\
    \hline
     by Danielle Walker, Senior Reporter July 22, 2014 Op Emmental spoofs bank sites, and uses Android malware to maintain account access So far, Operation Emmental has targeted victims' accounts in Austria, Switzerland, Sweden, and Japan. Cybercriminals have crafted an intricate attack method for maintaining a foothold in victim's online bank accounts, researchers warn. On Tuesday, Trend Micro released a 20-page report (PDF) on “Operation Emmental,” which makes use of Android malware capable of beating two-factor authentication and also delivers malicious code that changes an infected computer's DNS settings so it points to attacker-operated servers. Saboteurs begin the scheme by delivering malware through phishing attacks – malicious links or attachments designed to look like correspondence from popular retailers	
     & [Operation Emmental[ThreatActor], targets, Austria[Location]], [Operation Emmental[ThreatActor], targets, Switzerland[Location]], [Operation Emmental[ThreatActor], targets, Sweden[Location]], [Operation Emmental[ThreatActor], targets, Japan[Location]]
     & [Operation Emmental[ThreatActor], targets, Austria[Location]], [Operation Emmental[ThreatActor], targets, Switzerland[Location]], [Operation Emmental[ThreatActor], targets, Sweden[Location]], [Operation Emmental[ThreatActor], targets, Japan[Location]], [Operation Emmental[ThreatActor], uses, Android malware[MalwareType]], [Operation Emmental[ThreatActor], uses, malicious code[MalwareType]], [Operation Emmental[ThreatActor], uses, phishing attacks[AttackPattern]], [Operation Emmental[ThreatActor], uses, malware capable of beating two-factor authentication[AttackPattern]], [Operation Emmental[ThreatActor], discoveredIn, July 22, 2014[Time]], [Trend Micro[Organization], hasAuthor, Danielle Walker[Person]], [Trend Micro[Organization], hasAlias, Senior Reporter[Person]], [Trend Micro[Organization], indicates, Operation Emmental has targeted victims' accounts[AttackPattern]], [Trend Micro[Organization], discoveredIn, July 22, 2014[Time]]	\\
    \hline
\end{tabular}
\end{table*}

Further inductive link prediction with the NodePiece \cite{galkin2022nodepiece} implementation in \textit{pykeen} library \cite{ali2021pykeen} was conducted on an inference dataset containing entities not included in the test data. The results of the two models, one trained with the extracted triples from LADDER and one with triples from the fine-tuned 7B chat model, are very close. This implies that the relationships within the triples are fairly clear, enabling both models to effectively generalize from observed to unseen data. \textit{For example}, the trained NodePiece is capable of inferring a new relationship between the malware \code{Android/AdDisplay.Ashas}, which was not included in the training data, and the \code{Android} operating system. It accurately predicts this relationship when queried with the malware as the head entity and \code{targets} as the relationship. The second most likely prediction for this relationship is \code{Russian}, reflecting the significant number of users in Russia who were affected by this malware \cite{eset_ashas_2019}.

\begin{highlighted}{}
\footnotesize
\textbf{Finding 4: }The KG constructed from unstructured CTI using LLM exhibits promising link prediction capabilities, with models trained on extracted triples showing enhanced performance and robust generalization from known to unknown data relationships in certain scenarios.
\end{highlighted}

\section{Discussion}
\subsection{Case Study}

In this section, we present additional case studies for a qualitative assessment and comparison of fine-tuning and guidance methods. As shown in the outputs from both models in Table~\ref{tab:casestudy}, the fine-tuned 7B chat model confines its responses to malware-relevant triples and adheres to the specific ontology it was trained on during the fine-tuning process. On the other hand, the 70B chat model guided by prompts encounters difficulties in this respect. For instance in the first example, it incorrectly associates the entity type \code{MalwareType} with the relationships \code{targets}, \code{isA}, or \code{uses} in connection to \code{MalwareType} and \code{Person}, or assigns the type \code{Organization} the relationship \code{hasAuthor} with \code{MalwareType}. Additionally, it extracts triples related to the MITRE ATT\&CK framework mentioned in the paragraph, which are irrelevant to the malware discussion.

The same issue is apparent in the second example, where the model incorrectly extracts the irrelevant triple \code{[Trend Micro[Organization], hasAuthor, Danielle Walker[Person]]} and mistakenly links the type \code{Organization} with \code{indicates} to \code{AttackPattern}. However, in this case, the guidance model successfully extracts additional pertinent information, such as \code{[Operation Emmental[ThreatActor], uses, malware capable of bypassing two-factor authentication[AttackPattern]]}, which the fine-tuned model overlooks.

Despite the challenges, after filtering out triples that do not adhere to the ontology, the fine-tuned model ultimately produced a larger number of relevant triples, indicating that the guidance model's issues with relevance and accuracy have a more significant impact.

\subsection{Limitations and Future Work}

The study highlighted areas for potential improvement. 
Given the trial-and-error nature of prompt development and the vast range of possibilities, there is always the potential for discovering more effective prompts that could lead to enhanced model performance for guidance or the fine-tuning method. 
Additionally, the scope of annotated data available for fine-tuning was limited; access to a larger annotated dataset would likely enhance the model's accuracy, reduce bias, and improve generalizability in the fine-tuned model. Refining data pre-processing techniques could further improve results, particularly by removing clearly irrelevant data, as models still appear to struggle with distinguishing such content.
Moreover, the challenge of employing link prediction without attack pattern mapping was amplified by the extensive variety of potential connections, indicating areas for future enhancements.

\section{Conclusion}
This research demonstrates the effectiveness of using Large Language Models (LLMs) and Knowledge Graphs (KGs) for automating the extraction of actionable Cyber Threat Intelligence (CTI) from unstructured data. It highlights the impact of various extraction methodologies, finding that guidance framework significantly improved the output beyond what is achievable with few-shot prompt engineering alone. Furthermore, the study underlines the importance of fine-tuning for extracting triples that adhere to a specified ontology.
While the approach shows promise in structuring vast amounts of CTI and aiding in threat understanding, challenges remain in scaling and refining the methods for broader application. This study lays the groundwork for future advancements in automated CTI processing using modern technologies, indicating a significant leap forward in cybersecurity defenses.

\section*{Acknowledgments}

As part of the open-report model followed by the Workshop on Attackers \& CyberCrime Operations (WACCO), all the reviews for this paper are publicly available at \url{https://github.com/wacco-workshop/WACCO/tree/main/WACCO-2024}.

\bibliographystyle{IEEEtran}
\bibliography{references}

\begin{thebibliography}{10}
\providecommand{\url}[1]{#1}
\csname url@samestyle\endcsname
\providecommand{\newblock}{\relax}
\providecommand{\bibinfo}[2]{#2}
\providecommand{\BIBentrySTDinterwordspacing}{\spaceskip=0pt\relax}
\providecommand{\BIBentryALTinterwordstretchfactor}{4}
\providecommand{\BIBentryALTinterwordspacing}{\spaceskip=\fontdimen2\font plus
\BIBentryALTinterwordstretchfactor\fontdimen3\font minus \fontdimen4\font\relax}
\providecommand{\BIBforeignlanguage}[2]{{%
\expandafter\ifx\csname l@#1\endcsname\relax
\typeout{** WARNING: IEEEtran.bst: No hyphenation pattern has been}%
\typeout{** loaded for the language `#1'. Using the pattern for}%
\typeout{** the default language instead.}%
\else
\language=\csname l@#1\endcsname
\fi
#2}}
\providecommand{\BIBdecl}{\relax}
\BIBdecl

\bibitem{brown_sans_2023}
\BIBentryALTinterwordspacing
R.~Brown and K.~Nickels, ``{SANS 2023 CTI Survey: Keeping Up with a Changing Threat Landscape},'' 2023. [Online]. Available: \url{https://www.sans.org/white-papers/2023-cti-survey-keeping-up-changing-threat-landscape}
\BIBentrySTDinterwordspacing

\bibitem{rani_ttphunter_2023}
N.~Rani, B.~Saha, V.~Maurya, and S.~K. Shukla, ``{TTPHunter: Automated Extraction of Actionable Intelligence as TTPs from Narrative Threat Reports},'' in \emph{Australasian Information Security Conference (AISC 2023)}.\hskip 1em plus 0.5em minus 0.4em\relax {ACM}, 2023, pp. 126--134.

\bibitem{alam_looking_2023}
M.~T. Alam, D.~Bhusal, Y.~Park, and N.~Rastogi, ``{Looking Beyond IoCs: Automatically Extracting Attack Patterns from External CTI},'' in \emph{The 26th International Symposium on Research in Attacks, Intrusions and Defenses (RAID ’23)}, 2023.

\bibitem{wang_gpt-ner_2023}
\BIBentryALTinterwordspacing
{Wang et al.}, ``{GPT-NER: Named Entity Recognition via Large Language Models},'' 2023. [Online]. Available: \url{http://arxiv.org/abs/2304.10428}
\BIBentrySTDinterwordspacing

\bibitem{wan_gpt-re_2023}
\BIBentryALTinterwordspacing
{Wan et al.}, ``{GPT-RE: In-context Learning for Relation Extraction using Large Language Models},'' 2023. [Online]. Available: \url{http://arxiv.org/abs/2305.02105}
\BIBentrySTDinterwordspacing

\bibitem{wadhwa_revisiting_2023}
S.~Wadhwa, S.~Amir, and B.~Wallace, ``{Revisiting Relation Extraction in the era of Large Language Models},'' in \emph{Proceedings of the 61st Annual Meeting of the Association for Computational Linguistics}.\hskip 1em plus 0.5em minus 0.4em\relax Association for Computational Linguistics, 2023, pp. 15\,566--15\,589.

\bibitem{zhu_llms_2023}
\BIBentryALTinterwordspacing
{Zhu et al.}, ``{LLMs for Knowledge Graph Construction and Reasoning: Recent Capabilities and Future Opportunities},'' 2023. [Online]. Available: \url{http://arxiv.org/abs/2305.13168}
\BIBentrySTDinterwordspacing

\bibitem{han_pive_2023}
\BIBentryALTinterwordspacing
J.~Han, N.~Collier, W.~Buntine, and E.~Shareghi, ``{PiVe: Prompting with Iterative Verification Improving Graph-based Generative Capability of LLMs},'' 2023. [Online]. Available: \url{http://arxiv.org/abs/2305.12392}
\BIBentrySTDinterwordspacing

\bibitem{trajanoska_enhancing_2023}
\BIBentryALTinterwordspacing
M.~Trajanoska, R.~Stojanov, and D.~Trajanov, ``Enhancing knowledge graph construction using large language models,'' 2023. [Online]. Available: \url{http://arxiv.org/abs/2305.04676}
\BIBentrySTDinterwordspacing

\bibitem{fayyazi_uses_2023}
\BIBentryALTinterwordspacing
R.~Fayyazi and S.~J. Yang, ``{On the Uses of Large Language Models to Interpret Ambiguous Cyberattack Descriptions}.'' [Online]. Available: \url{http://arxiv.org/abs/2306.14062}
\BIBentrySTDinterwordspacing

\bibitem{juttner_chatids_2023}
\BIBentryALTinterwordspacing
V.~Jüttner, M.~Grimmer, and E.~Buchmann, ``{ChatIDS: Explainable Cybersecurity Using Generative AI},'' 2023. [Online]. Available: \url{http://arxiv.org/abs/2306.14504}
\BIBentrySTDinterwordspacing

\bibitem{gupta_chatgpt_2023}
\BIBentryALTinterwordspacing
M.~Gupta, C.~Akiri, K.~Aryal, E.~Parker, and L.~Praharaj, ``{From ChatGPT to ThreatGPT: Impact of Generative AI in Cybersecurity and Privacy},'' 2023. [Online]. Available: \url{http://arxiv.org/abs/2307.00691}
\BIBentrySTDinterwordspacing

\bibitem{touvron_llama_2023}
\BIBentryALTinterwordspacing
{Touvron et al.}, ``Llama 2: Open foundation and fine-tuned chat models,'' 2023. [Online]. Available: \url{http://arxiv.org/abs/2307.09288}
\BIBentrySTDinterwordspacing

\bibitem{jiang_mistral_2023}
\BIBentryALTinterwordspacing
{Jiang et al.}, ``Mistral 7b,'' 2023. [Online]. Available: \url{http://arxiv.org/abs/2310.06825}
\BIBentrySTDinterwordspacing

\bibitem{tunstall_zephyr_2023}
\BIBentryALTinterwordspacing
{Tunstall et al.}, ``Zephyr: Direct distillation of {LM} alignment,'' 2023. [Online]. Available: \url{http://arxiv.org/abs/2310.16944}
\BIBentrySTDinterwordspacing

\bibitem{openai_prompt}
\BIBentryALTinterwordspacing
{OpenAI}, ``{Prompt engineering - OpenAI Platform}.'' [Online]. Available: \url{https://platform.openai.com/docs/guides/prompt-engineering}
\BIBentrySTDinterwordspacing

\bibitem{noauthor_guidance-aiguidance_2024}
\BIBentryALTinterwordspacing
S.~Lundberg, H.~Nori, and M.~T. Ribeiro, ``guidance-ai/guidance.'' [Online]. Available: \url{https://github.com/guidance-ai/guidance}
\BIBentrySTDinterwordspacing

\bibitem{hu_lora_2021}
\BIBentryALTinterwordspacing
{Hu et al.}, ``{LoRA}: Low-rank adaptation of large language models,'' 2021. [Online]. Available: \url{http://arxiv.org/abs/2106.09685}
\BIBentrySTDinterwordspacing

\bibitem{johnson_guide_2016}
\BIBentryALTinterwordspacing
C.~S. Johnson, M.~L. Badger, D.~A. Waltermire, J.~Snyder, and C.~Skorupka, ``{Guide to Cyber Threat Information Sharing},'' 2016, {NIST} {SP} 800-150. [Online]. Available: \url{https://nvlpubs.nist.gov/nistpubs/SpecialPublications/NIST.SP.800-150.pdf}
\BIBentrySTDinterwordspacing

\bibitem{symantec}
\BIBentryALTinterwordspacing
{Symantec}, ``{Symantec Enterprise Blogs/Threat Intelligence},'' 2024. [Online]. Available: \url{https://symantec-enterprise-blogs.security.com/blogs/threat-intelligence}
\BIBentrySTDinterwordspacing

\bibitem{mandiant}
\BIBentryALTinterwordspacing
Mandiant, ``Threat intelligence reports,'' 2024. [Online]. Available: \url{https://www.mandiant.com/resources/reports}
\BIBentrySTDinterwordspacing

\bibitem{crowdstrike}
\BIBentryALTinterwordspacing
{CrowdStrike}, ``{CrowdStrike Blog},'' 2024. [Online]. Available: \url{https://www.crowdstrike.com/blog}
\BIBentrySTDinterwordspacing

\bibitem{dalziel2014define}
H.~Dalziel, \emph{{How to define and build an effective cyber threat intelligence capability}}.\hskip 1em plus 0.5em minus 0.4em\relax Syngress, 2014.

\bibitem{husari_ttpdrill_2017}
G.~Husari, E.~Al-Shaer, M.~Ahmed, B.~Chu, and X.~Niu, ``{TTPDrill: Automatic and Accurate Extraction of Threat Actions from Unstructured Text of CTI Sources},'' in \emph{Proceedings of the 33rd Annual Computer Security Applications Conference}.\hskip 1em plus 0.5em minus 0.4em\relax {ACM}, 2017, pp. 103--115.

\bibitem{zhu_chainsmith_2018}
Z.~Zhu and T.~Dumitras, ``{ChainSmith: Automatically Learning the Semantics of Malicious Campaigns by Mining Threat Intelligence Reports},'' in \emph{2018 {IEEE} European Symposium on Security and Privacy ({EuroS}\&P)}.\hskip 1em plus 0.5em minus 0.4em\relax {IEEE}, 2018, pp. 458--472.

\bibitem{satyapanich_casie_2020}
T.~Satyapanich, F.~Ferraro, and T.~Finin, ``{CASIE: Extracting Cybersecurity Event Information from Text},'' in \emph{Proceedings of the {AAAI} Conference on Artificial Intelligence}, vol.~34, no.~5, 2020, pp. 8749--8757.

\bibitem{aghaei_securebert_2022}
\BIBentryALTinterwordspacing
E.~Aghaei, X.~Niu, W.~Shadid, and E.~Al-Shaer, ``{SecureBERT}: A domain-specific language model for cybersecurity,'' 2022. [Online]. Available: \url{http://arxiv.org/abs/2204.02685}
\BIBentrySTDinterwordspacing

\bibitem{devlin2019bert}
\BIBentryALTinterwordspacing
J.~Devlin, M.-W. Chang, K.~Lee, and K.~Toutanova, ``{BERT: Pre-training of Deep Bidirectional Transformers for Language Understanding},'' 2019. [Online]. Available: \url{https://arxiv.org/abs/1810.04805}
\BIBentrySTDinterwordspacing

\bibitem{liu2019roberta}
\BIBentryALTinterwordspacing
Y.~Liu, M.~Ott, N.~Goyal, J.~Du, M.~Joshi, D.~Chen, O.~Levy, M.~Lewis, L.~Zettlemoyer, and V.~Stoyanov, ``{RoBERTa: A Robustly Optimized BERT Pretraining Approach},'' 2019. [Online]. Available: \url{https://arxiv.org/pdf/1907.11692}
\BIBentrySTDinterwordspacing

\bibitem{vaswani_attention_2017}
{Vaswani et al.}, ``{Attention is All you Need},'' in \emph{31st Conference on Neural Information Processing Systems (NIPS 2017), Long Beach, CA, USA.}, 2017.

\bibitem{zhao_survey_2023}
\BIBentryALTinterwordspacing
{Zhao et al.}, ``{A Survey of Large Language Models},'' 2023. [Online]. Available: \url{http://arxiv.org/abs/2303.18223}
\BIBentrySTDinterwordspacing

\bibitem{dettmers_qlora_2023}
\BIBentryALTinterwordspacing
T.~Dettmers, A.~Pagnoni, A.~Holtzman, and L.~Zettlemoyer, ``{QLoRA: Efficient Finetuning of Quantized LLMs},'' 2023. [Online]. Available: \url{http://arxiv.org/abs/2305.14314}
\BIBentrySTDinterwordspacing

\bibitem{noauthor_llm_nodate}
\BIBentryALTinterwordspacing
{Hugging Face}. {LLM prompting guide}. [Online]. Available: \url{https://huggingface.co/docs/transformers/main/tasks/prompting}
\BIBentrySTDinterwordspacing

\bibitem{mitre_web}
\BIBentryALTinterwordspacing
{The MITRE Corporation}, ``{MITRE ATT\&CK®},'' 2024. [Online]. Available: \url{https://attack.mitre.org}
\BIBentrySTDinterwordspacing

\bibitem{li_instruction-tuned_nodate}
X.~Li, F.~Polat, and P.~Groth, ``{Do Instruction-tuned Large Language Models Help with Relation Extraction?}'' in \emph{KBC-LM’23: Knowledge Base Construction from Pre-trained Language Models workshop at ISWC 2023}, 2023.

\bibitem{zhang_aligning_2023}
\BIBentryALTinterwordspacing
K.~Zhang, B.~J. Gutiérrez, and Y.~Su, ``{Aligning Instruction Tasks Unlocks Large Language Models as Zero-Shot Relation Extractors},'' 2023. [Online]. Available: \url{http://arxiv.org/abs/2305.11159}
\BIBentrySTDinterwordspacing

\bibitem{siracusano_time_2023}
\BIBentryALTinterwordspacing
{Siracusano et al.}, ``{Time for aCTIon: Automated Analysis of Cyber Threat Intelligence in the Wild},'' 2023. [Online]. Available: \url{http://arxiv.org/abs/2307.10214}
\BIBentrySTDinterwordspacing

\bibitem{rossi_knowledge_2021}
A.~Rossi, D.~Firmani, A.~Matinata, P.~Merialdo, and D.~Barbosa, ``{Knowledge Graph Embedding for Link Prediction: A Comparative Analysis},'' in \emph{{ACM} Transactions on Knowledge Discovery from Data}, 2021, vol.~15, no.~2, pp. 1--49.

\bibitem{ghidini_sepses_2019}
E.~Kiesling, A.~Ekelhart, K.~Kurniawan, and F.~Ekaputra, ``{The SEPSES Knowledge Graph: An Integrated Resource for Cybersecurity},'' in \emph{The Semantic Web – {ISWC} 2019}.\hskip 1em plus 0.5em minus 0.4em\relax Springer International Publishing, 2019, vol. 11779, pp. 198--214.

\bibitem{sun_aptkg_2022}
L.~Sun, Z.~Li, L.~Xie, M.~Ye, and B.~Chen, ``{APTKG: Constructing Threat Intelligence Knowledge Graph from Open-Source APT Reports Based on Deep Learning},'' in \emph{2022 5th International Conference on Data Science and Information Technology ({DSIT})}.\hskip 1em plus 0.5em minus 0.4em\relax {IEEE}, 2022, pp. 01--06.

\bibitem{li_attackg_2022}
\BIBentryALTinterwordspacing
Z.~Li, J.~Zeng, Y.~Chen, and Z.~Liang, ``{AttacKG: Constructing Technique Knowledge Graph from Cyber Threat Intelligence Reports},'' 2022. [Online]. Available: \url{http://arxiv.org/abs/2111.07093}
\BIBentrySTDinterwordspacing

\bibitem{gao_threatkg_2022}
\BIBentryALTinterwordspacing
{Gao et al.}, ``{ThreatKG: A Threat Knowledge Graph for Automated Open-Source Cyber Threat Intelligence Gathering and Management},'' 2022. [Online]. Available: \url{http://arxiv.org/abs/2212.10388}
\BIBentrySTDinterwordspacing

\bibitem{rastogi_tinker_2023}
N.~Rastogi, S.~Dutta, M.~J. Zaki, A.~Gittens, and C.~Aggarwal, ``{TINKER: A framework for Open source Cyberthreat Intelligence},'' in \emph{2022 IEEE International Conference on Trust, Security and Privacy in Computing and Communications (TrustCom)}, 2022.

\bibitem{rotate_2019}
\BIBentryALTinterwordspacing
Z.~Sun, Z.~Deng, J.~Nie, and J.~Tang, ``{RotatE: Knowledge Graph Embedding by Relational Rotation in Complex Space},'' \emph{CoRR}, vol. abs/1902.10197, 2019. [Online]. Available: \url{http://arxiv.org/abs/1902.10197}
\BIBentrySTDinterwordspacing

\bibitem{Jiang_convr_2019}
X.~Jiang, Q.~Wang, and B.~Wang, ``{Adaptive Convolution for Multi-Relational Learning},'' in \emph{North American Chapter of the Association for Computational Linguistics}, 2019.

\bibitem{balazevic_tucker_2019}
I.~Balazevic, C.~Allen, and T.~Hospedales, ``{TuckER: Tensor Factorization for Knowledge Graph Completion},'' in \emph{Proceedings of the 2019 Conference on Empirical Methods in Natural Language Processing and the 9th International Joint Conference on Natural Language Processing ({EMNLP}-{IJCNLP})}.\hskip 1em plus 0.5em minus 0.4em\relax Association for Computational Linguistics, 2019, pp. 5184--5193.

\bibitem{ali2021improving}
\BIBentryALTinterwordspacing
{Ali et al.}, ``{Improving Inductive Link Prediction Using Hyper-Relational Facts},'' 2021. [Online]. Available: \url{https://arxiv.org/abs/2107.04894}
\BIBentrySTDinterwordspacing

\bibitem{galkin2022nodepiece}
\BIBentryALTinterwordspacing
M.~Galkin, E.~Denis, J.~Wu, and W.~L. Hamilton, ``{NodePiece: Compositional and Parameter-Efficient Representations of Large Knowledge Graphs},'' 2022. [Online]. Available: \url{https://arxiv.org/abs/2106.12144}
\BIBentrySTDinterwordspacing

\bibitem{liu2022review}
\BIBentryALTinterwordspacing
K.~Liu, F.~Wang, Z.~Ding, S.~Liang, Z.~Yu, and Y.~Zhou, ``{A review of knowledge graph application scenarios in cyber security},'' 2022. [Online]. Available: \url{https://arxiv.org/abs/2204.04769}
\BIBentrySTDinterwordspacing

\bibitem{openai_bestpractices}
\BIBentryALTinterwordspacing
{OpenAI}, ``Best practices for prompt engineering with {OpenAI} {API} {\textbar} {OpenAI} help center.'' [Online]. Available: \url{https://help.openai.com/en/articles/6654000-best-practices-for-prompt-engineering-with-openai-api}
\BIBentrySTDinterwordspacing

\bibitem{rechia_deep_2023}
\BIBentryALTinterwordspacing
P.~Rechia, ``{A Deep Dive Into Guidance’s Source Code},'' 2023. [Online]. Available: \url{https://betterprogramming.pub/a-deep-dive-into-guidances-source-code-16681a76fb20}
\BIBentrySTDinterwordspacing

\bibitem{alpaca}
{Taori et al.}, ``{Stanford Alpaca: An Instruction-following LLaMA model},'' \url{https://github.com/tatsu-lab/stanford_alpaca}, 2023.

\bibitem{hf_howtogenerate}
\BIBentryALTinterwordspacing
P.~von Platen, ``{How to generate text: using different decoding methods for language generation with Transformers}.'' [Online]. Available: \url{https://huggingface.co/blog/how-to-generate}
\BIBentrySTDinterwordspacing

\bibitem{hf_textgenerationstrategy}
\BIBentryALTinterwordspacing
{Hugging Face}, ``Text generation strategies.'' [Online]. Available: \url{https://huggingface.co/docs/transformers/main/generation_strategies}
\BIBentrySTDinterwordspacing

\bibitem{huggingface}
\BIBentryALTinterwordspacing
------, ``{Hugging Face Models},'' 2024. [Online]. Available: \url{https://huggingface.co/models}
\BIBentrySTDinterwordspacing

\bibitem{transformers}
\BIBentryALTinterwordspacing
------, ``{Hugging Face Transformers},'' 2024. [Online]. Available: \url{https://github.com/huggingface/transformers}
\BIBentrySTDinterwordspacing

\bibitem{bitsandbytes}
\BIBentryALTinterwordspacing
T.~Dettmers, ``{bitsandbytes},'' 2024. [Online]. Available: \url{https://github.com/TimDettmers/bitsandbytes}
\BIBentrySTDinterwordspacing

\bibitem{autogptq}
\BIBentryALTinterwordspacing
{AutoGPTQ}, ``{AutoGPTQ},'' 2024. [Online]. Available: \url{https://github.com/AutoGPTQ/AutoGPTQ}
\BIBentrySTDinterwordspacing

\bibitem{peft}
\BIBentryALTinterwordspacing
{Hugging Face}, ``{peft},'' 2024. [Online]. Available: \url{https://github.com/huggingface/peft}
\BIBentrySTDinterwordspacing

\bibitem{trl}
\BIBentryALTinterwordspacing
------, ``{trl},'' 2024. [Online]. Available: \url{https://github.com/huggingface/trl}
\BIBentrySTDinterwordspacing

\bibitem{accelerate}
\BIBentryALTinterwordspacing
------, ``{accelerate},'' 2024. [Online]. Available: \url{https://github.com/huggingface/accelerate}
\BIBentrySTDinterwordspacing

\bibitem{lin_rouge_nodate}
C.-Y. Lin, ``{ROUGE: A Package for Automatic Evaluation of Summaries},'' in \emph{Annual Meeting of the Association for Computational Linguistics}, 2004.

\bibitem{ali2021pykeen}
M.~Ali, M.~Berrendorf, C.~T. Hoyt, L.~Vermue, S.~Sharifzadeh, V.~Tresp, and J.~Lehmann, ``{PyKEEN 1.0: A Python Library for Training and Evaluating Knowledge Graph Embeddings},'' \emph{Journal of Machine Learning Research}, vol.~22, no.~82, pp. 1--6, 2021.

\bibitem{eset_ashas_2019}
\BIBentryALTinterwordspacing
L.~Stefanko, ``{ESET Research: Tracking down the developer of Android adware affecting millions of users},'' 2019. [Online]. Available: \url{https://www.welivesecurity.com/2019/10/24/tracking-down-developer-android-adware/}
\BIBentrySTDinterwordspacing

\end{thebibliography}

\end{document}